\documentclass[useAMS,usenatbib]{mn2e}

\usepackage[dvips]{graphicx}
\usepackage[dvips]{graphicx}
\usepackage{dcolumn}
\usepackage{times}
\title[Image reconstruction technique
 and optical monitoring of the QSO2237+0305]{Image reconstruction technique
 and optical monitoring of the QSO2237+0305 from Maidanak Observatory in 2002 -- 2003}
\author[E.~Koptelova et al.]{E.~Koptelova$^{1}$\thanks{E-mail:
koptelova@xray.sai.msu.ru}, E.~Shimanovskaya$^{1}$\thanks{E-mail:
eshim@sai.msu.ru}, B.~Artamonov$^{1}$\thanks{E-mail:
artamon@sai.msu.ru}, M.~Sazhin$^{1}$, A.~Yagola$^{2}$,
\newauthor V.~Bruevich$^{1}$ and O.~Burkhonov$^{3}$\\
$^{1}$Sternberg Astronomical Institute, Universitetski pr. 13, 119992 Moscow, Russia\\
$^{2}$Moscow State University, Faculty of Physics, Vorobiovi
Gori, 119992 Moscow, Russia\\
$^{3}$Ulugh Beg Astronomical Institute of Ac.Sci. of Uzbekistan,
Astronomicheskaya 33, 700052 Tashkent, Uzbekistan Republic}
\begin{document}

\date{26 August 2004}

\pagerange{\pageref{firstpage}--\pageref{lastpage}} \pubyear{2004}

\maketitle

\label{firstpage}

\begin{abstract}

We have observed the gravitational lens system Q2237+0305 from the
Maidanak Observatory over the period from August 2002 to November
2003. Here we report the results of our observations. We
implemented a two-stage technique that has been developed
specifically for the purpose of gravitational lens image
reconstruction. The technique is based on the Tikhonov
regularization approach and allows one to obtain astrometric and
photometric characteristics of the gravitational lens system.
Light curves with 78 data points for the four quasar components
are obtained. Slow brightness variations over the observational
period are found in all components. Images A, C, D have a
tendency to decrease in brightness. Image B does not vary more
than 0.05mag. The observations did not reveal evidence for large
variations in brightness of the components due to microlensing
effects. To provide an overall picture of the photometry
behaviour, our data are combined with the Maidanak observations
published for 1995 -- 2000.
\end{abstract}


\section{Introduction}

Since 1997, an international program of gravitational lens
monitoring has been carried out at Maidanak Observatory
(Uzbekistan) by Tashkent, Moscow and Kharkov observational
groups. For the period 1997 -- 2003 a huge set of observational
data was gathered for the following well known gravitational lens
systems: Q2237+0305 (Einstein Cross), SBS1520+530, SBS0909+532,
PG1115+080, H1413+117, RX1413+117, RX0921+4528, UM673
(QSO0142-100), B1422+231. The aim of the monitoring program is to
obtain light curves for these lens systems.

We have developed a photometric method for the reliable treatment
of gravitational lenses with a visible lensing galaxy and applied
it to the 2 year dataset of Q2237+0305.

Here, we present results for Q2237+0305, a rather complex system
which is the object of monitoring programs of many observational
groups. It consists of a barred spiral galaxy at a redshift
$z_{\rm d}$=0.039, in which \citet{huchra1985} discovered a
high-redshift quasar ($z_{\rm s}$=1.695). The location of the
lensed images, which are close to the bulge of the lensing
galaxy, makes microlensing events highly probable in this system
\citep{kayser1989}. The regular observations of Q2237+0305 began
from the first microlensing event observed in August 1989 by
\citet{irwin1989}. The first attempt to get the light curves of
the four quasar components was made by \citet{corrigan1991}. The
similar attempt was made in 1994 through constructing light
curves which were free from the effects of different spectral
bands \citep{houde1994}. The long-duration monitoring program of
Q2237+0305 was started in 1990 at the Nordic Optical Telescope
\citep{ostensen1996}. During this five-year monitoring program
microlensing variations had been detected in all four images. The
most densely sampled four-year monitoring was conducted within
the OGLE program from August 1997 to November 2000
\citep{vozniak2000}. It revealed brightness variations in all
images and range up to 1.2mag for image C in 1999. The GLITP
collaboration presented the results for the period from October
1999 to February 2000 \citep{alcalde2002}. The GLITP data, which
covered a time-period of greater than one month, began after a
high magnification microlensing event was observed for the A
component by the OGLE collaboration. The monitoring observations
from Apache Point Observatory \citep{schmidt2002} presented the
light curves for only the A and B components of Q2237+0305 from
June 1995 to January 1998, but they contained the brightness peak
of A component in 1996. The results of VRI photometry in 1997 --
1998 from Maidanak Observatory were published by
\citet{bliokh1999} and \citet{dudinov2000}. The combined VRI
light curves from the monitoring program at the Maidanak
Observatory in 1995 -- 2000 were presented by \citet{dudinov2004}.

In this paper we present R-band observations of Q2237+0305 from
August 2002 to November 2003. The next section describes the
conditions of the observations followed by the description of the
two-stage image reconstruction technique. We then show the results
of the image reconstruction over the observational period and
present a photometric variability plot for the quasar components.

\section{Observations}
Observations of Q2237+0305 were carried out with the 1.5-m AZT-22
telescope of the high-altitude Maidanak Observatory (Uzbekistan)
using the LN-cooled CCD camera of Copenhagen University
Observatory with the imaging area of 2000$\times$800 and a pixel
size of 15$\mu$m, giving a spatial sampling 0.26 arcsec
pixel$^{-1}$. Data were taken in r the Gunn filter, which
corresponds approximately to the standard Johnson-Cousins system.
The poor tracking available at the telescope only allowed
exposures up to a maximum of 3 minutes. To obtain sufficiently
high photometric accuracy with such short exposures, the images
were taken in series, consisting of 4--8 frames each. The seeing
conditions are presented in
Tables~\ref{photometry02}~and~\ref{photometry03} via the values
of FWHM for particular nights. The best quality of the image
corresponds to the point source with FWHM=0.75\arcsec.
Preprocessing of the data (including bias-level subtraction,
flat-field division, sky subtraction and cosmic ray removal) was
done with the standard routines in the Munich Image Data Analysis
System (MIDAS) environment. Several stars in the imaging area were
used as reference stars to reduce all frames to the same
coordinate system. To increase signal-to-noise ratio, and 'to
reveal' underlying galaxy several images with excellent seeing
(FWHM$\simeq0.9\arcsec$) were summed before being subjected to
photometric processing. Then a subframe of 64 by 64 pixels
centred on the nucleus of the galaxy 2237+0305 was extracted.

\section{Two-stage image \\* reconstruction algorithm}

An essential prerequisite for accurate photometry in ground-based
images is good seeing well below the source separation. The
extremely compact spatial structure of Q2237+0305, where the
component separation is comparable to the seeing, complicates
accurate photometry. The presence of the lensing galaxy, with a
point-like nucleus and an extended light distribution, makes the
photometric results dependant on the galaxy model. This
peculiarity leads to poor agreement between the results of
different monitoring programs, which adopt different ways to
extract the underlying galaxy flux. These difficulties have been
noted by many authors
\citep{yee1988,corrigan1991,vakulik1997,burud1998}. In
minimization procedures, the galaxy brightness distribution is
usually represented either analytically
\citep{ostensen1996,alcalde2002,teuber1993} or numerically
\citep{burud1998,magain1998}. To solve this problem, iterative
algorithms are often used to approximately realize minimization
of the $\chi^{2}$-function. These algorithms estimate the flux
contribution from the underlying galaxy, which can then be used
in the photometry of the lensed images. Such an approach
noticeably simplifies the solution procedure, and provides good
intrinsic convergence \citep{corrigan1991,alcalde2002,burud1998}
but unfortunately does not ensure the absence of systematic
errors in estimating the magnitudes of the components caused by a
poor galaxy model. For the photometry of the data we used a
two-stage algorithm developed for the image reconstruction of the
objects with point sources superimposed on a smooth background.
The algorithm enables the complex images to be split up into the
numerical lensing galaxy and quasar components, and allows the
galaxy to be subtracted from the individual images which are
under consideration. To optimize the photometric treatment of
observational data, the process is divided into two stages: 1)
combining several images with good seeing and extracting a
numerical galaxy model from the combined frame with a
regularizing algorithm; 2) processing large numbers of image
frames using this numerical galaxy model to get astrometric and
photometric characteristics of the gravitational lens system.

{\it The first stage.} The first stage of the algorithm is based
on the Tikhonov regularization approach. Images obtained on
ground-based telescopes affected by the finite instrument
resolution and atmospheric perturbations. The simple model of the
image formation can be represented as a convolution equation:
\begin{equation}
\label{fredholm} (t*z)(x,y)=\int\!\!\!\int\limits_{\!\! \rm B}
t(x-\xi,y-\eta)z(\xi,\eta)\,d\xi d\eta=u (x,y),
\end{equation}
where $z(x,y)$ is the unknown light distribution of the object, $u
(x,y)$ represents the observable light distribution, the kernel of
the above equation $t(x,y)$ is the point spread function (PSF).
The estimate of the PSF can be obtained from the images of
reference stars in the neighbourhood of the object. In this work
two approaches were used to determine the PSF. One approach
involves adopting a reference star profile as a numerical point
spread function. This method is free of any assumptions about the
shape of the PSF, but it depends on the star location because the
PSF can vary over the frame field. Another approach involves the
theoretical elliptical Gaussian distribution. We also used the
semi-analytical RAS (Rotate-And-Stare) method described by
\citet{teuber1994} and employed in the XECClean package by
\citet{ostensen1996}. PSFs determined with the RAS method are in a
good agreement with those theoretically modelled by elliptical
Gaussian distributions. The PSF is constructed 1) using an
individual star ($\alpha$) and 2) using the mean brightness
distribution obtained via superposition of profiles of reference
stars ($\alpha, \beta$) weighted inversely proportional to their
intensities. The terminology of \citet{yee1988} and
\citet{corrigan1991} was adopted for the reference stars. We found
that the best model for the PSF was obtained using the $\alpha$
star.

The observable data are registered on the pixel grid with some
error. The presence of photon noise and, as a consequence, the
error of the input data complicates the problem (\ref{fredholm})
which otherwise can be solved in Fourier space. So, the problem
of image reconstruction lies in finding the approximate solution
of the equation (\ref{fredholm}) with the approximate kernel $t$
with known error estimation {\it h}:
\begin{equation}
\label{hh} \sup_{\|z\|_{\rm Z}=1}\|t\ast z-\bar{t}\ast z\|_{\rm
L_{2}}\leq h,
\end{equation}
and having at our disposal the noisy data $u$ and the estimate of
the noise level: \[\sigma_{\rm tot}: \|u-\bar{u}\|_{\rm
L_2}\leq\sigma_{\rm tot}.\] Here, barred letters $\bar{t}$ and
$\bar{u}$ denote exact, non-error contaminated, PSF and observed
image respectively. $\|\cdot\|_{\rm Z}$ denotes the norm in the
$\rm Z$ set of functions. It is assumed that the observed image
$u$ belongs to the space $\rm L_2$ of square integrable functions
with the norm:
\begin{equation}
\label{uu} \|u\|_{\rm L_2}\equiv \int\!\!\!\int\limits_{\!\! \rm
B}u^2(x,y)dxdy
\end{equation}
The estimation of the $\sigma_{\rm tot}$ can be calculated as the
total noise integrated over the resulting subframe:
\begin {equation}
\sigma_{\rm
tot}=\sqrt{\sum_{i,j}(\frac{I_{ij}}{g}+{N}\cdot{R^{2}})},
\label{noise}
\end {equation}
where {\it N} is the number of summed frames with the same
exposure time; {\it I$_{ij}$} -- counts in the {\it ij}-pixel of
the subframe; $g$ -- gain factor of the CCD camera; {\it R} --
readout noise. The error of the operator (\ref{hh}) depends on the
method of PSF modeling. If the relative error of PSF modeling is
$d$, then the error of definition of the operator can be
estimated as follows:
\begin{equation}
h=\sqrt{\sum_{i,j} (d \cdot t_{ij})^{2}}.
\end{equation}

To reduce the systematic errors connected with the galaxy model
we try to construct the perfect galaxy model applying
regularization techniques. The convolution equation
(\ref{fredholm}) with the error contaminated right-hand side
belongs to the class of ill-posed inverse problems. The solution
may be non-unique, and small variations in the input data may
lead to large variations in the solution. To solve such problems
a regularization method was developed by A. Tikhonov
\citep{tikhars,tikhonov} that allows to find approximate solution
taking into account {\it a priori} information about the
structure of the object. This unique solution, which possesses a
specified degree of smoothness and provides a physical
representation of the lensing galaxy's flux distribution, tends
to the true solution in the norm of the functional space chosen
when the errors in input data tend to zero.

\begin{figure*}
%\begin{center}
\vspace*{150pt}
\includegraphics[scale=1,bb=30 5 460 10,angle=0]{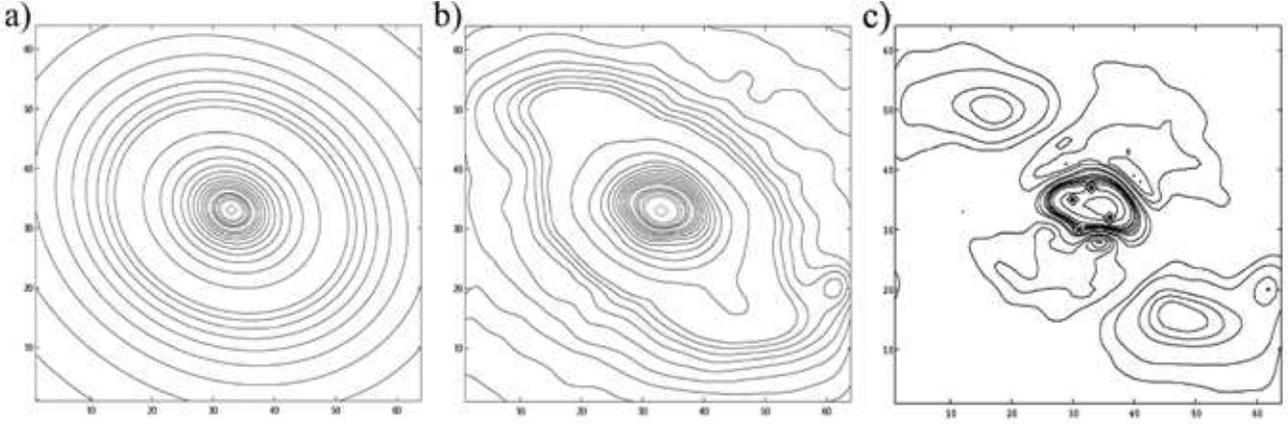}
  \caption{The contours of the central region ($17.14\arcsec \times 17.14\arcsec$) of the lensing galaxy:
   a) the analytical galaxy model (Sersic model fitting); b) the numerical galaxy model (Tikhonov regularization algorithm); c) the difference between a) and b) with the quasar components superimposed.}
  \label{profiles}
%\end{center}
\end{figure*}

The regularization requires construction of an algorithm to
control the trade-off between (a) the assumptions about both the
smoothness and the structure of the sought solution and (b) its
consistency with the data. The key concept of the algorithm is a
smoothing function of the following kind:
\begin{equation}
\label{regular} M^\alpha[z]=\|t*z-u\|^{2}_{\rm
L_2}+\alpha\cdot\Omega[z]
\end{equation}
Here the first term represents the squared discrepancy between the
model and data, $\alpha$ is the regularization parameter which
controls the balance between the consistency of the sought
solution with the right side of (\ref{fredholm}) and the
smoothness of the sought solution, $\Omega[z]$ is a stabilizer
function through which {\it a priori} information is introduced
into the problem formulation \citep{yagola2003}. Let $z^{\alpha}$
be the extremum of the function $M^\alpha[z]$ on $Z$, i.e.
$z^{\alpha}$ is the solution of the minimization problem for
$M^\alpha[z]$ on the chosen set of functions (possibly with some
constraints).

The choice of the regularization parameter $\alpha$, which
provides uniform residuals and the necessary smoothness of the
solution, is crucial for solving ill-posed problems. Generally,
it should depend on the input data, the errors, and the method of
approximation of the initial problem. One of the way to
co-ordinate the regularization parameter with the error of the
input information is the discrepancy principle -- adoption of
$\alpha>0$ satisfying:
\begin{equation}
\label{discprinc} \|t*z^{\alpha}-u\|_{\rm L_2}\simeq\sigma_{\rm
tot}.
\end{equation}
The regularization parameter $\alpha$ is chosen such that the
corresponding residual (the left hand side of (\ref{discprinc}))
is equal to the a priori specified bound (right hand side) for the
noise level in the image. The regularization method with $\alpha$
chosen according to the discrepancy principle (\ref{discprinc})
is convergent and of optimal order \citep{morozov1984,engl2000}.
Provided that the regularization parameter $\alpha$ is chosen
according to this rule, the solution  $z^{\alpha}$ of the
minimization problem for $M^{\alpha}[z]$ can be considered as an
approximate solution. The approximate solution obtained using
this method goes to the true solution when the error of input
data goes to zero.

When the kernel $t$ in (\ref{fredholm}) is known inexactly, one
can use the generalized discrepancy principle
\citep{nonlinearillposed} which lies in solving the following
equation:
\begin{eqnarray}
\rho(\alpha)=0 \label{nev}
\end{eqnarray}
where $\rho(\alpha)$~is called the generalized discrepancy:
\begin{eqnarray}
\rho(\alpha)\equiv \|t*z^{\alpha}-u\|_{\rm L_{2}}^{2}-
\left(\sigma_{\rm tot} + h \sqrt{\Omega[z^{\alpha}}]\right)^{2}
\label{nevdef}.
\end{eqnarray}

The function $\rho(\alpha)$ is strictly monotonic and the root of
(\ref{nev}) can be found by bisection or some other standard
method. This is an {\it a posteriori} method of choosing the
regularization parameter, because, to find the root of equation
(\ref{nev}), one needs to minimize the smoothing function at
every iteration by $\alpha$. Given the estimations of
$\sigma_{\rm tot}$ and $h$ we used the generalized discrepancy
principle to choose the regularization parameter.

Various assumptions about the structure of the object under study
can also be taken into account. Images of close quadruple
gravitational lens systems consist of multiple overlapped quasar
images superimposed on a lensing galaxy. So, the image can be
decomposed into two constituent parts: the sum of four
$\delta$-functions and smooth background (galaxy). Written for
the pixel grid, that assumption is represented as follows:
\begin{equation}
z_{mn}=\sum_{q=1}^{4} a_{q} \delta_{m-b_{q},n-c_{q}}+g_{mn}
\label{split},
\end{equation}
where $a_{q}$ are the intensities of point sources with
coordinates $(b_{q},c_{q})$, $\delta_{m-b_{q},n-c_{q}}$ are
Kronnecker deltas, $g_{mn}$ is the solution's component
corresponding to the lensing galaxy.

Numerical simulations revealed that introducing the assumption
about the closeness of the real galaxy light distribution to some
analytical profile produces more stable results for the
reconstruction. In this work we assume that the light
distribution in the central region of the galaxy is well-modelled
by a generalized de Vaucouleurs profile
\citep{devauc1948,devauc1959}, known as Sersic's model
\citep{sersic1968,cardone2004}:
\begin{equation}
g^{\rm mod}(r)=I_{\rm e}\exp\left({-b_{n} (\frac{r}{r_{\rm
e}})^{\frac{1}{n}}}\right) \label{gal},
\end{equation}
where $b_{n} = 2n - 0.324$ for $1 < n < 4$ ($n=4$ corresponds to
de Vaucouleurs model), $r_{\rm e}$ is an effective radius, $I_{\rm
e}$ is an intensity within the effective radius, and $r(x_{\rm
c},y_{\rm c},a_{\rm x},a_{\rm y},\theta)$:
\begin{equation}
r^2=\frac{x'^2}{a_{\rm x}^2}+\frac{y'^2}{a_{\rm y}^2}
\end{equation}
where $x'$ and $y'$ are measured from the centre along the major
and minor axes.

Given the image, one can represent it in accordance with
(\ref{split}) where the lensing galaxy is described with the
analytical model (\ref{gal}). Parameters of the analytical model
are fitted by means of minimization of the $\chi^2$-function:
\[
\chi^2(a_{q},b_{q},c_{q},I_{\rm e},r_{\rm e},n,x_{\rm c},y_{\rm
c},a_{\rm x},a_{\rm y},\theta)=
\]%\nonumber\\
\begin{equation}
=\sum_{i,j}^{N_1 N_2}\frac{1}{\sigma_{ij}^{2}}( \sum_{m,n}^{N_1
N_2} t_{i-m,j-n} \cdot z_{mn}^{\rm mod}-u_{ij})^{2}  ,
%\label{hi},
\end{equation}
\[
z_{mn}^{\rm
mod}=\sum_{q=1}^{4}a_{q}\delta_{m-b_{q},n-c_{q}}+g_{mn}^{\rm mod}
,
\]
where $\sigma_{ij}$ is the noise in each pixel of the frame. The
positions of quasar components $(b_q,c_q)$, the corresponding
amplitudes $a_q$, and the analytical galaxy model parameters are
fitted. Powell \citep{press1998} method is used as a minimization
routine.

Given the assumption that the real galaxy light distribution is
close to the analytical profile the stabilizer function in
(\ref{regular}) can be written as follows:
\begin{eqnarray}
\Omega[z] = \|g^{\rm num}-g^{\rm mod}\|_{\rm G}^{2} + \beta
\sum_{q=1}^{4} a_{q}^{2} \label{stab},
\end{eqnarray}
where $g^{\rm num}$ is a sought numerical galaxy model, $g^{\rm
mod}$ is the analytical galaxy model. The second term in the
right-hand side is introduced in order to penalize values of
quasar intensities that are too large compared to the lensing
galaxy. The parameter $\beta$ is chosen on the basis of the model
calculations in the way that both terms in the stabilizer function
are of the same order. Such an approach enables probable
artefacts in the galaxy brightness distribution (`holes' at the
positions of quasar components) to be excluded.

After the parameters of the analytical galaxy model $g^{\rm mod}$
have been found, this model is used for construction of the
stabilizer function and as an initial approach for minimization
of the smoothing function:
\[
M^{\alpha}(a_q,g^{\rm num})= \sum_{i,j}^{N_1
N_2}\frac{1}{\sigma_{ij}^{2}}\left( \sum_{m,n}^{N_1 N_2}
t_{i-m,j-n} z_{mn} -u_{ij}\right)^{2}+
\]
\begin{equation}
+\alpha \sum_{i,j}^{N_1 N_2}\left( g_{ij}^{\rm num}-g_{ij}^{\rm
mod}\right)^2+ \beta \sum_{q=1}^{4}a_{q}    , \label{mdiscr}
\end{equation}
\[
z_{mn}=\sum_{q=1}^{4}a_{q}\delta_{m-\bar{b}_{q},n-\bar{c}_{q}}+g^{\rm
num}_{mn}.
\]

Here $(\bar{b}_q,\bar{c}_q)$ are the fixed quasar components
coordinates from the preliminary step. The method of conjugate
gradients was used to construct the minimizing sequence
\citep{press1998}.

Due to the huge number of free parameters the process of
minimizing the smoothing function is time-consuming. The main
result at this stage is the numerical galaxy model. After the
accurate numerical galaxy model is found, the huge data set for
the same observational period can be processed in a much shorter
space of time. The equal brightness contours for the central part
of the galaxy for both analytical and numerical galaxy models are
presented in Fig.~\ref{profiles}. The mean values of the galaxy
parameters from the CASTLE collaboration in H filter, the GLITP
collaboration, and our values derived from the first stage of the
two-stage image reconstruction algorithm in R filter are combined
in Table~\ref{galaxypar}.

\begin{table}
\begin{center}
\caption{Parameters of the lensing galaxy: $r_{\rm e}$ --
effective radius, $\epsilon$ -- ellipticity, P.A. -- position
angle.}
\begin{tabular}{cr@{$\pm$}lr@{$\pm$}lr@{$\pm$}lr@{$\pm$}l} \hline
images& \multicolumn{2}{c}{$r_{\rm e}$ (\arcsec)}&
\multicolumn{2}{c}{$\epsilon$} & \multicolumn{2}{c}{P.A.
($\circ$)}
\\ \hline
H CASTLES & 4.7& 0.9 & 0.33&0.01 &66&1   \\
R GLITP & 4.94& 0.25 & 0.38&0.02 &62&1   \\
R Maidanak & 4.6& 0.3 & 0.35&0.02 &64&1   \\
\hline
\end{tabular}
\label{galaxypar}
\end{center}
\end{table}

{\it The second stage.} In the second stage, the numerical galaxy
model $g^{\rm num}$, obtained in the previous stage as a result
of the minimization of the Tikhonov regularization function
(\ref{mdiscr}), is used for the photometric treatment of all
observational data. For every individual frame the galaxy
brightness distribution is described as follows:
\begin{eqnarray}
G_{mn}=\lambda_{1}\cdot g^{\rm num}_{mn}+\lambda_{2} \label{gala},
\end{eqnarray}
where $\lambda_{1}$ is a multiplier giving a level of the galaxy
intensity in every individual frame, $\lambda_{2}$ is some
additional constant background. So, since a galaxy in the second
stage is described only by two parameters, the number of
parameters are reduced allowing a decrease in processing time.
The finite-dimensional solution of this fitting problem can be
found by minimizing a $\chi^{2}$-function of the following form:
\[
\chi^2(a_{q},b_{q},c_{q},\lambda_{1},\lambda_{2})=
\]
%\begin{equation}
%\nonumber\\
\[\sum_{i,j}^{N_1 N_2}\frac{1}{\sigma_{ij}^{2}}(
\sum_{m,n}^{N_1 N_2}\{t_{i-m,j-n}
(\sum_{q=1}^{4}a_{q}\delta_{m-b_{q},n-c_{q}}+
%\]
%\[+
G_{mn})\}-u_{ij})^{2} \label{hisec}.
\]
%\end{equation}
%
As far as exposure time for all processed frames is the same, the
overall brightness of the galaxy in the subframe should be nearly
constant. Calculations exhibited that the parameter $\lambda_{1}$
fluctuates by 1.4\% from subframe to subframe. That slight
variation might be due to different seeing conditions.

The results of the image reconstruction from the second stage are
presented in Fig.~\ref{recon}.

\section{Results}

\begin{figure*}
\begin{center}
\vspace*{110pt}
\includegraphics[scale=1.0, bb=5 -5 450 30,angle=0]{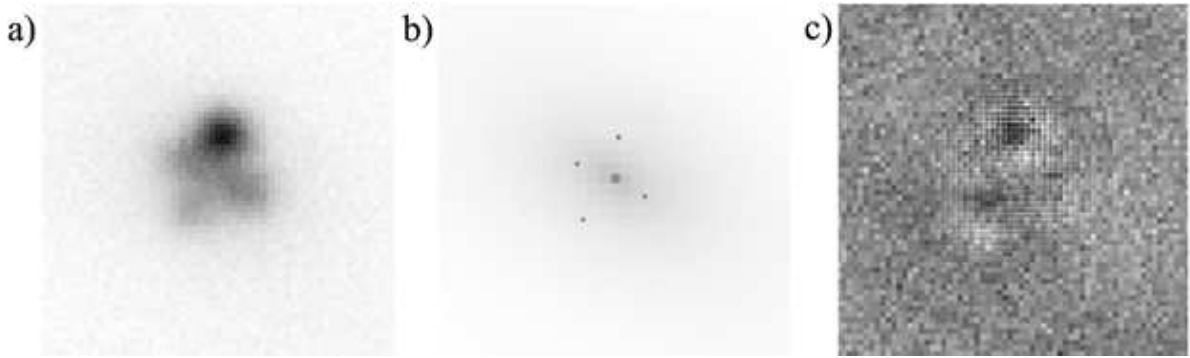}
\caption{Results of the image reconstruction of gravitational lens
system Q2237+0305 (Einstein Cross) with our two-stage algorithm:
a) observed image ($OBS$); b) reconstructed image ($REC$); c)
residuals ($\frac{OBS-convolved REC}{\sqrt{convolved REC}}$)}
\label{recon}
\end{center}
\end{figure*}

We have developed a two-stage technique for the image
reconstruction of objects with a complex structure that consist
of point sources superimposed on a smooth lensing galaxy. The
algorithm that we have developed has been shown to perform well in
reconstructing images of a quadruple gravitational lens system.
We believe it is flexible enough to be applied to other objects
with structure of a similar complexity. The algorithm allows the
adoption of different analytical models to 'adjust' a numerical
galaxy model and also allows the choice of different sets of
functions to control the smoothness of the background constituent
(lensing galaxy) of the solution.

The technique was used to process the two-year observational data
for the close quadruple gravitational lens system Q2237+0305, well
known for its complex structure and the small separation between
quasar components. The correct photometry for such kind of objects
is not possible without constructing as realistic a model for the
underlying galaxy as possible. We try to get an accurate galaxy
model by avoiding an analytical description of the galaxy. This
ensures that the numerical galaxy model obtained in the first
stage of the algorithm is free from any disadvantages associated
with an analytical model. Fig.~\ref{profiles} shows contours of
analytical (a) and numerical (b) galaxy models constructed in the
first stage of the algorithm. The contour map of the difference
between the analytical and numerical galaxy models in the Fig.
\ref{profiles}c) reveals an uncertainty in the central region
where the quasar components are located, and in the spiral barred
regions. Fig.~\ref{profiles}b) shows that the numerical galaxy
has a realistic structure with arms. It is evident that the
residual map in Fig.~\ref{recon}c) does not contain any
significant contribution from the spiral arms of the galaxy,
unlike the analytical galaxy modelling approach. The numbers of
quasar image positions are presented in the Table~\ref{qsopos}
along with CASTLES data.

\begin{table}
\begin{center}
\caption{Astrometry of the components of Q2237+0305 relative to A
component.}
\begin{tabular}{cr@{$\pm$}lr@{$\pm$}lr@{$\pm$}lr@{$\pm$}lr@{$\pm$}lr@{$\pm$}l} \hline
{} & \multicolumn{4}{c}{H CASTLES} & \multicolumn{4}{c}{R
Maidanak}\\
{}& \multicolumn{2}{c}{$\Delta$R.A. (\arcsec)} &
\multicolumn{2}{c}{$\Delta$Dec. (\arcsec)} &\multicolumn{2}{c}
{$\Delta$R.A. (\arcsec)} & \multicolumn{2}{c}{$\Delta$Dec.
(\arcsec)}
\\ \hline
B & -0.673 & 0.003 & 1.697&0.003 & -0.723 & 0.052 & 1.646 & 0.058 \\
C & 0.635  & 0.003 & 1.209&0.003 &  0.567 & 0.069 & 1.211 & 0.016 \\
D & -0.866 & 0.003 & 0.528&0.003 & -0.858 & 0.016 & 0.534 & 0.016 \\
G & -0.075 & 0.003 & 0.939&0.003 & -0.094 & 0.028 & 0.937 & 0.027 \\
\hline
\end{tabular}
\label{qsopos}
\end{center}
\end{table}

To evaluate the ability of the algorithm to fit the observational
data, the $\chi^2$ per degree of freedom ($\overline{\chi}^2$ )
was used. In this problem the number of degrees of freedom is
equal 4108, corresponding to the pixel size of the frame which is
under the treatment ($64\times 64$ pixels) plus two coordinates
and the flux of each quasar component. The reduced
$\overline{\chi}^{2}$ value for the two-stage image
reconstruction method varied in the range of 1.2 to 9.0.

The compactness of the system (approximately 2 arcsec) further
complicates the photometric treatment. Unfortunately, despite the
intrinsically good seeing, a poor telescope tracking system
significantly reduces the quality of the data and in some cases
induces ellipticity in the PSF. In our data processing we use two
strategies for the PSF construction. In cases where the image
quality was compromised by the poor tracking system, the
numerical PSF seems to be more suitable.

We present monitoring data of Q2237+0305 over the period from 28
August 2002 to 27 November 2003. The R-band light curves for the
four images are shown in Fig. \ref{lightcurves0203} and tabulated
in Tables~\ref{photometry02}~and~\ref{photometry03}. Since
Maidanak data have been taken in Gunn r-band, the reference
$\alpha$ star with known r-magnitude determined by
\citet{corrigan1991} was used for the calibration of the
Q2237+0305 data. Transition to the standard system was carried
out according to the color equation with coefficients derived
from the Maidanak observations of the Landolt standards fields
\citep{landolt1992}. The color correction was done using the
results from \citet{vakulik2003}. The quoted error values in
Tables~\ref{photometry02}~and~\ref{photometry03} are the standard
deviation of the processed frames for one night. Note that these
errors do not include any error associated with the method, since
this is impossible to calculate for ill-posed problems.

\section{Discussion}

\begin{figure*}
\begin{center}
\vspace*{290pt}
\includegraphics[scale=1., bb=350 5 200 70,angle=0]{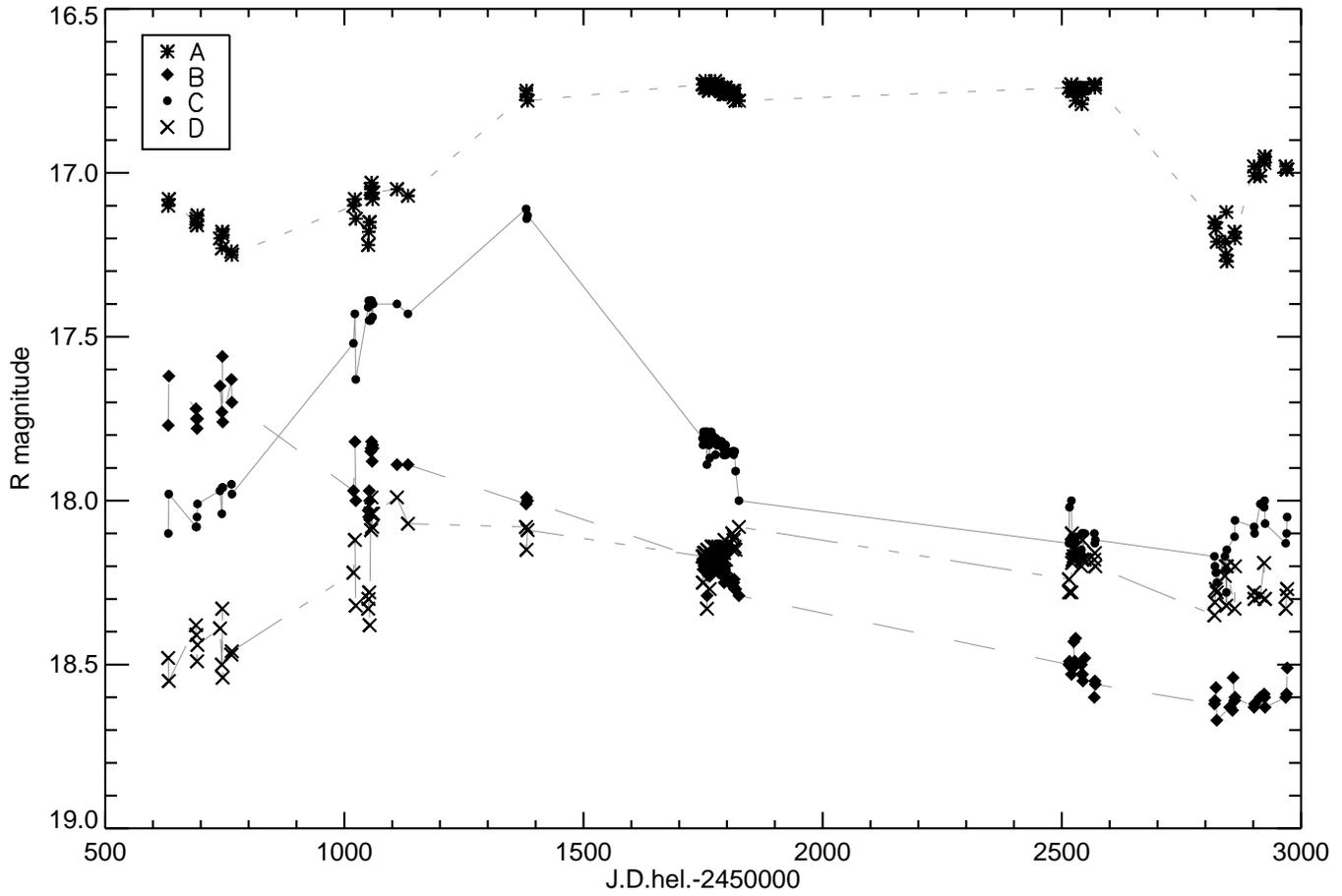}
  \caption{Combined Q2237+0305 light curves in the R band for 1995 -- 2003
based on data from \citet{vakulik2003} and from this paper.}
  \label{combinedlc}
\end{center}
\end{figure*}

In Fig. \ref{combinedlc} we have combined the R-band light curves
for the four quasar components of Q2237+0305 over the period from
1995 to 2003, based on data obtained in 1995 -- 2000 by
\citet{vakulik2003} and on the results of the photometry
treatment for 2002 -- 2003 presented in this paper. All plotted
data correspond to observations conducted at the Maidanak
Observatory.

As can be seen, trends in the light curves of the C and D
components are fairly similar. The other two components show less
agreement. There is a drop for the A component in the July 2003
data points, but there is a rise of 0.2mag at the end of the light
curve. Future behaviour of this component could be detected with
further observations. For the B component the light curve is
flatter than for the others. As the B component of the system is
the faintest one, some discrepancies for several days and the
large scatter in the data points may be due to days with bad
seeing.

It is useful to compare the presented results with ones obtained
for the same observational period with another photometry method.
To test our two-stage image reconstruction algorithm, we processed
several images from the Maidanak dataset with the CLEAN procedure
\citep{ostensen1996} which revealed that for the images with
seeing 0.8-0.9 arcsec the results of the photometry for the A
component match those obtained using the proposed technique.
Getting accurate quasar image positions, which maximize the
correlation coefficients of the CLEAN method, from images with
worse seeing is rather problematic, considering that correlation
coefficients less than 0.98 lead to unacceptable large
uncertainty in the magnitude values.

\begin{figure*}
%\begin{center}
\vspace*{190pt}
\includegraphics[scale=1, bb=20 5 510 160,angle=0]{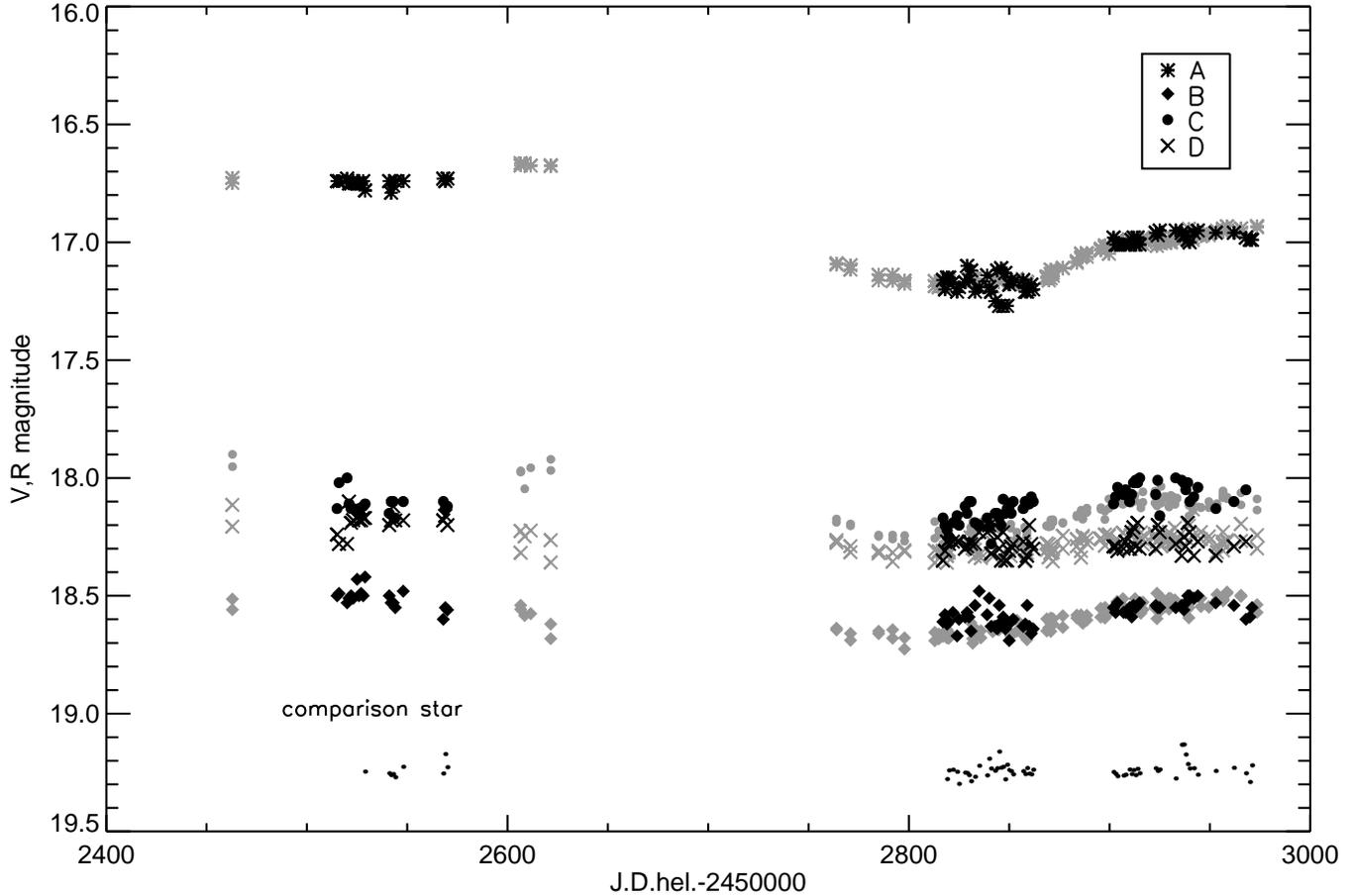}
\caption{Photometry of Q2237+0305 in R band from observations
with 1.5-m Maidanak telescope in 2002-2003. Photometry in V band
by OGLE program
(http://www.astrouw.edu.pl/~ogle/ogle3/huchra.html) is also
plotted by fainter symbols. The comparison star
 ($\alpha$) is shifted in magnitude by 2.7mag.}
\label{lightcurves0203}
%\end{center}
\end{figure*}

Another test for the algorithm is the comparison of the results
with light curves obtained by OGLE monitoring program
(http://www.astrouw.edu.pl/~ogle/ogle3/huchra.html).
Fig.~\ref{lightcurves0203} shows a good agreement between the
results of two-stage algorithm and photometry of Q2237+0305 in V
band by OGLE program.

\section*{Acknowledgments}

The authors would like to thank the referee for a very helpful
report.

We gratefully acknowledge the use of data obtained by the
German-Uzbek collaboration between Potsdam University (Robert
Schmidt, Joachim Wambsganss), Astrophysical Institute Potsdam
(Stefan Gottl\"ober, Lutz Wisotzki) and the Ulugh Beg
Astronomical Institute Tashkent (Salakhutdin Nuritdinov).

This research was supported by Russian Foundation for Basic
Research (RFBR) grants 02-01-00044, and 01-02-16800. We thank
Alexandr Gusev for providing transformation coefficients of the
Maidanak instrumental system to the standard photometric system,
and Vasily Belokurov for detail discussion of this work and
helpful suggestions. We thank Andrzej  Udalski for the attention
to our work, and Martin Smith for the careful reading of the
paper.

\begin{table*}
\begin{center}
\caption{Photometry of Q2237+0305 in R filter from observations at
Maidanak Observatory in 2002. The table contains the date in
yymmdd format, the Julian date (-2450000), the seeing as
determined from $\alpha$ star and the magnitudes of four quasar
components.}
\begin{tabular}{cccr@{$\pm$}lr@{$\pm$}lr@{$\pm$}lr@{$\pm$}l} \hline
date& Julian date & seeing (arcsec) & \multicolumn{2}{c}{A}&
\multicolumn{2}{c}{B} & \multicolumn{2}{c}{C} &
\multicolumn{2}{c}{D}
\\ \hline
02.08.28 & 2515& 0.8 & 16.74&0.01 &18.62&0.02 & 18.21&0.01&18.23&0.02  \\
02.08.29 & 2516& 0.9 & 16.74&0.04 &18.49&0.04 & 18.02&0.04&18.38&0.06 \\
02.09.02 & 2520& 1.3 & 16.73&0.03 &18.53&0.07 & 18.00&0.09&18.28&0.09 \\
02.09.03 & 2521& 1.2 & 16.75&0.01 &18.51&0.05 & 18.11&0.07&18.10&0.09 \\
02.09.04 & 2522& 0.8 & 16.75&0.01 &18.50&0.10 & 18.13&0.03&18.19&0.03 \\
02.09.05 & 2523& 0.7 & 16.74&0.02 &18.51&0.09 & 18.13&0.02&18.18&0.09 \\
02.09.07 & 2525& 0.8 & 16.75&0.07 &18.43&0.02 & 18.14&0.03&18.17&0.05 \\
02.09.08 & 2526& 0.9 & 16.75&0.02 &18.50&0.02 & 18.13&0.05&18.18&0.05 \\
02.09.09 & 2527& 0.8 & 16.75&0.03 &18.49&0.07 & 18.16&0.09&18.18&0.04  \\
02.09.10 & 2528& 0.8 & 16.74&0.01 &18.50&0.02 & 18.12&0.03&18.17&0.02  \\
02.09.17 & 2529& 1.1 & 16.78&0.07 &18.42&0.10 & 18.11&0.03&18.10&0.10  \\
02.09.23 & 2541& 1.1 & 16.74&0.07 &18.50&0.08 & 18.15&0.05&18.10&0.20  \\
02.09.24 & 2542& 1.0 & 16.79&0.05 &18.53&0.09 & 18.10&0.04&18.18&0.01  \\
02.09.25 & 2543& 1.0 & 16.76&0.03 &18.53&0.05 & 18.15&0.04&18.12&0.09  \\
02.09.26 & 2544& 1.3 & 16.74&0.07 &18.64&0.19 & 18.17&0.13&18.16&0.09  \\
02.09.30 & 2548& 1.0 & 16.74&0.05 &18.48&0.10 & 18.10&0.12&18.18&0.09  \\
02.10.20 & 2568& 0.9 & 16.73&0.04 &18.60&0.10 & 18.10&0.05&18.18&0.07  \\
02.10.21 & 2569& 1.0 & 16.74&0.02 &18.55&0.08 & 18.13&0.09&18.16&0.03  \\
02.10.22 & 2570& 0.9 & 16.73&0.03 &18.56&0.07 & 18.12&0.07&18.20&0.05  \\
\hline
\end{tabular}
\label{photometry02}
\end{center}
\end{table*}

\begin{table*}
\begin{center}
\caption{Photometry of Q2237+0305 in R filter from observations at
Maidanak Observatory in 2003. The table contains the date in
yymmdd format, the Julian date (-2450000), the seeing as
determined from $\alpha$ star and the magnitudes of four quasar
components.}
%\newcolumntype{d}{D{\pm}{\pm}{-1}}
%\begin{tabular}{cr@{$\pm$}lr@{$\pm$}lr@{$\pm$}lr@{$\pm$}lr@{$\pm$}lr@{$\pm$}l}
%\hline

\begin{tabular}{cccr@{$\pm$}lr@{$\pm$}lr@{$\pm$}lr@{$\pm$}l} \hline
date& Julian date & seeing (arcsec) & \multicolumn{2}{c}{A}&
\multicolumn{2}{c}{B} & \multicolumn{2}{c}{C} &
\multicolumn{2}{c}{D}
\\ \hline

03.06.26 & 2817& 1.3 & 17.16&0.06 &18.61&0.10 & 18.21&0.11&18.35&0.08  \\
03.06.27 & 2818& 1.0 & 17.20&0.09 &18.58&0.10 & 18.18&0.08&18.32&0.11  \\
03.06.28 & 2819& 1.1 & 17.15&0.04 &18.62&0.09 & 18.17&0.11&18.35&0.17  \\
03.06.29 & 2820& 0.9 & 17.15&0.05 &18.61&0.09 & 18.20&0.07&18.31&0.11  \\
03.07.01 & 2822& 1.0 & 17.17&0.07 &18.57&0.01 & 18.22&0.09&18.27&0.11  \\
03.07.03 & 2824& 1.7 & 17.21&0.12 &18.67&0.11 & 18.25&0.15&18.28&0.15  \\
03.07.04 & 2825& 1.0 & 17.19&0.07 &18.60&0.10 & 18.19&0.04&18.27&0.13  \\
03.07.07 & 2828& 0.9 & 17.17&0.04 &18.59&0.09 & 18.16&0.07&18.27&0.10  \\
03.07.08 & 2829& 1.0 & 17.10&0.02 &18.57&0.08 & 18.20&0.13&18.27&0.09  \\
03.07.09 & 2830& 1.5 & 17.16&0.07 &18.59&0.10 & 18.12&0.11&18.30&0.12  \\
03.07.10 & 2831& 1.5 & 17.12&0.05 &18.65&0.12 & 18.15&0.11&18.29&0.13  \\
03.07.12 & 2833& 1.4 & 17.21&0.07 &18.54&0.09 & 18.10&0.08&18.27&0.11  \\
03.07.14 & 2835& 1.1 & 17.19&0.06 &18.48&0.12 & 18.10&0.06&18.28&0.08  \\
03.07.18 & 2839& 1.2 & 17.14&0.09 &18.58&0.09 & 18.19&0.11&18.28&0.12  \\
03.07.19 & 2840& 1.0 & 17.19&0.06 &18.51&0.09 & 18.20&0.09&18.23&0.11  \\
03.07.20 & 2841& 1.2 & 17.21&0.03 &18.63&0.04 & 18.17&0.09&18.23&0.03  \\
03.07.22 & 2843& 1.4 & 17.25&0.07 &18.63&0.10 & 18.21&0.11&18.20&0.07  \\
03.07.23 & 2844& 1.3 & 17.12&0.09 &18.64&0.10 & 18.28&0.11&18.32&0.21  \\
03.07.24 & 2845& 1.3 & 17.27&0.09 &18.54&0.09 & 18.15&0.05&18.20&0.18  \\
03.07.25 & 2846& 1.2 & 17.11&0.06 &18.61&0.10 & 18.15&0.09&18.25&0.08  \\
03.07.26 & 2847& 1.1 & 17.27&0.09 &18.59&0.09 & 18.17&0.05&18.30&0.11  \\
03.07.27 & 2848& 1.4 & 17.13&0.09 &18.64&0.09 & 18.20&0.08&18.35&0.18  \\
03.07.28 & 2849& 1.2 & 17.27&0.09 &18.61&0.06 & 18.09&0.06&18.23&0.10  \\
03.07.29 & 2850& 1.2 & 17.18&0.07 &18.69&0.09 & 18.15&0.04&18.35&0.15  \\
03.07.30 & 2851& 1.2 & 17.16&0.08 &18.61&0.09 & 18.13&0.10&18.35&0.14  \\
03.07.31 & 2852& 0.9 & 17.17&0.04 &18.60&0.09 & 18.12&0.07&18.31&0.13  \\
03.08.05 & 2857& 1.2 & 17.16&0.09 &18.63&0.10 & 18.15&0.07&18.29&0.07  \\
03.08.06 & 2858& 0.9 & 17.21&0.05 &18.62&0.09 & 18.10&0.07&18.28&0.10  \\
03.08.07 & 2859& 0.9 & 17.21&0.08 &18.54&0.03 & 18.13&0.06&18.27&0.17  \\
03.08.08 & 2860& 0.9 & 17.20&0.04 &18.63&0.10 & 18.10&0.08&18.35&0.11  \\
03.08.09 & 2861& 0.8 & 17.18&0.04 &18.66&0.09 & 18.11&0.08&18.33&0.08  \\
03.08.10 & 2862& 0.9 & 17.20&0.03 &18.64&0.09 & 18.11&0.04&18.20&0.17  \\
03.09.19 & 2902& 1.1 & 16.98&0.07 &18.55&0.07 & 18.07&0.11&18.28&0.12  \\
03.09.20 & 2903& 0.8 & 17.01&0.05 &18.57&0.08 & 18.10&0.09&18.30&0.04  \\
03.09.21 & 2904& 1.3 & 17.01&0.08 &18.53&0.08 & 18.11&0.11&18.29&0.18  \\
03.09.24 & 2907& 1.1 & 17.01&0.03 &18.57&0.07 & 18.08&0.09&18.29&0.09  \\
03.09.25 & 2908& 1.0 & 17.01&0.04 &18.57&0.07 & 18.04&0.05&18.31&0.11  \\
03.09.27 & 2910& 1.3 & 17.01&0.04 &18.55&0.09 & 18.07&0.05&18.30&0.13  \\
03.09.28 & 2911& 1.4 & 16.99&0.09 &18.59&0.11 & 18.05&0.11&18.27&0.19  \\
03.09.29 & 2912& 1.2 & 16.98&0.09 &18.54&0.09 & 18.10&0.10&18.30&0.15  \\
03.09.30 & 2913& 0.8 & 17.01&0.03 &18.55&0.08 & 18.07&0.06&18.23&0.08  \\
03.10.01 & 2914& 0.9 & 16.98&0.06 &18.54&0.07 & 18.02&0.04&18.21&0.13  \\
03.10.02 & 2915& 0.8 & 17.01&0.05 &18.53&0.07 & 18.01&0.09&18.29&0.07  \\
03.10.10 & 2923& 0.9 & 16.96&0.02 &18.54&0.02 & 18.02&0.07&18.19&0.09  \\
03.10.11 & 2924& 1.0 & 16.97&0.02 &18.55&0.03 & 18.00&0.07&18.30&0.10  \\
03.10.12 & 2925& 1.1 & 16.95&0.02 &18.55&0.05 & 18.07&0.08&18.30&0.18  \\
03.10.20 & 2933& 1.0 & 16.95&0.02 &18.55&0.05 & 18.01&0.03&18.20&0.12  \\
03.10.23 & 2936& 1.4 & 16.96&0.03 &18.54&0.06 & 18.16&0.04&18.23&0.15  \\
03.10.24 & 2937& 1.1 & 16.97&0.03 &18.56&0.04 & 18.00&0.07&18.28&0.09  \\
03.10.25 & 2938& 0.9 & 16.97&0.03 &18.53&0.07 & 18.01&0.07&18.33&0.13  \\
03.10.26 & 2939& 1.1 & 16.99&0.05 &18.50&0.06 & 18.02&0.06&18.25&0.05  \\
03.10.27 & 2940& 1.4 & 17.00&0.05 &18.50&0.07 & 18.05&0.09&18.29&0.11  \\
03.10.29 & 2942& 1.0 & 16.96&0.05 &18.51&0.03 & 18.02&0.08&18.19&0.13  \\
03.10.31 & 2944& 1.2 & 16.95&0.03 &18.50&0.03 & 18.10&0.08&18.23&0.18  \\
03.11.09 & 2953& 1.2 & 16.96&0.02 &18.53&0.06 & 18.08&0.06&18.33&0.14  \\
03.11.18 & 2962& 0.9 & 16.96&0.02 &18.54&0.05 & 18.04&0.03&18.27&0.09  \\
03.11.24 & 2968& 1.1 & 16.98&0.05 &18.60&0.08 & 18.13&0.05&18.33&0.11  \\
03.11.26 & 2970& 1.4 & 16.99&0.10 &18.59&0.10 & 18.10&0.08&18.29&0.20  \\
03.11.27 & 2971& 0.9 & 16.99&0.04 &18.55&0.05 & 18.05&0.07&18.27&0.09  \\
\hline
\end{tabular}
\label{photometry03}
\end{center}
\end{table*}

\end{document}